\documentstyle[epsfig]{elsart}

\begin{document}
\hspace*{\fill}%
\begin{minipage}{7cm}
\flushright
preprint UMIST-TP-96/1\\
hep-ph/yymmxxx
\end{minipage}

\begin{frontmatter}
\title{Quantising the $B=2$ and $B=3$ Skyrmion systems}

\author{Niels R.\ Walet\thanksref{email}}
\thanks[email]{electronic address: Niels.Walet@umist.ac.uk}

\address{
Department of Physics, UMIST, P.O. Box 88,
Manchester M60 1QD, U.K.}

\begin{abstract}
We examine the quantisation of a collective Hamiltonian for the
two-baryon system derived by us in a previous paper. We show that
by increasing the sophistication of the approximations we can
obtain a bound state - or a resonance - not too far removed from
the threshold with the quantum numbers of the deuteron.
 The energy of this state is shown to depend
very sensitively on the parameters of the model.
Subsequently we construct part of a collective Hamiltonian for
the three baryon system. Large-amplitude quantum fluctuations play
an important r\^ole in the intrinsic wave function of the ground-state,
changing its symmetry  from octahedral to cubic. 
Apart from the tetrahedron describing the minimum of the potential,
we identify a ``doughnut'' and a ``pretzel''
as the most important saddle points in the potential energy surface.
We show that it is likely that inclusion of fluctuations through
these saddle points lead to an energy close to the triton's value.
\end{abstract}
\end{frontmatter}

\section{Introduction}\label{sec:Skyrme}
In a recent paper \cite{SkLACM} we have investigated a numerical procedure
to determine a collective coordinate manifold for the Skyrme model, and have
exhibited many of the parameters in the collective Hamiltonian. 
At almost the same time Leese, Manton and Schroers \cite{LeeseMantonSchroers} 
published a paper studying
the quantisation of the Skyrme model in the attractive channel. Even though
their collective Hamiltonian was not determined self-consistently, they
obtained some very interesting results. They found a bound state at 6 MeV
binding energy, with the quantum numbers of the deuteron. More
importantly they found a pion
density remarkably similar to that in the deuteron.

In this paper we concentrate on a description of the deuteron using
our collective Hamiltonian. We have a lot of additional information,
that can be used to construct several approximations of increasing
sophistication to the dynamics of the Skyrme model in the $12$-dimensional
manifold thought to describe the dynamics of the deuteron. We shall also
exhibit the dependence on the choice of parameters in the Skyrme model,
and shall argue that the result of Manton and collaborators was accidental,
and that the best one can hope for is an energy not too far from zero,
maybe with an error of 30 MeV or so. 

\begin{figure}[htb]
\epsfysize=6cm
\centerline{\epsffile{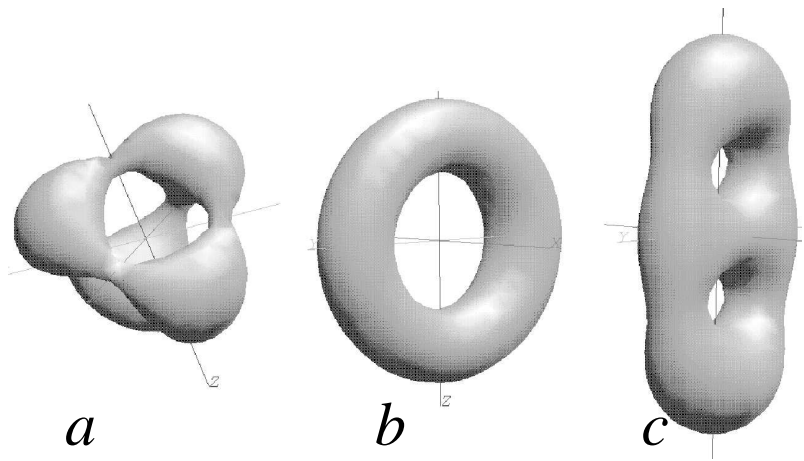}}
\caption{A plot of a surface of constant baryon density in (a) the
tetrahedron , (b) the $B=3$ doughnut and (c) the pretzel.}
\label{fig:Bdens}
\end{figure}

After having analysed the deuteron in great detail we turn our attention to
the triton and $^3$He. A while ago,
Carlson \cite{Carlson} has studied the application of
 the Skyrme model to the system of three baryons
(see also Ref.~\cite{LeeseManton} where the $B=3$ system is studied
using the approximate Yang-Mills instanton-induced (YMI-induced) 
form for the Skyrme fields).
In these studies one
 starts from the minimum energy solution where the baryon density
has tetrahedral symmetry (see Fig.~\ref{fig:Bdens}a). One then makes
the approximation that this minimum describes the triton. This
approximation leads to a tremendous over-binding, which was attributed
to the neglect of simple quantum effects, especially vibrational 
zero-point motion. It has also been mentioned \cite{Walhout} 
that anharmonic modes might play a r\^ole.

In this paper we show that the low-energy potential ``landscape'' has
a lot of structure, which has both important qualitative and quantitative 
effects. 
The most salient features in this landscape, 
apart from the tetrahedral solutions,
are two meta-stable states at slightly higher energy,
the $B=3$ doughnut and the ``pretzel'' which has a planar symmetry 
similar to the doughnut but has two holes
(see Figs.~\ref{fig:Bdens}b and c).

The approach taken is this work is based on the techniques discussed in
great detail in Ref.~\cite{SkLACM}, where we show how to study
large amplitude collective motion in the Skyrme model, using the YMI-induced
forms of the fields for simplicity.
This is based on a mode-following approach, where we start
from the harmonic fluctuations around a stable solution, and follow
those into the anharmonic regime. One advantage of our approach
is that such anharmonic modes always run through the extrema of the
potential energy. The main limitation of the current work is the
use of the instanton induced form of the Skyrme fields, 
see also Ref.~\cite{LeeseManton}. 

The paper is organised as follows. In Sec.~\ref{sec:skyrme} we introduce
the Skyrme model, and discuss the instanton-induced approximation
to the dynamics of the model. 
We also discuss the relation of the
scaled Skyrme units to standard units. 
We then discuss, in Sec.~\ref{sec:harmonic},
 how the study of harmonic fluctuations can be used to calculate
improved observables. In the next section, Sec.~\ref{sec:B=1},
 we succinctly recapitulate the
properties of the $B=1$ hedgehog within the YMI-induced Ansatz.
We then discuss the quantisation of the $B=2$ system in Sec.~\ref{sec:B=2}.
In Sec.~\ref{sec:LACMB=3} we
discuss the structure of the potential landscape, and how the
different structures are connected. In the next section (Sec.~\ref{sec:quantB=3})
we try to estimate what the effects of the structure of the potential
landscape are on the ground-state of the triton.
Finally, we draw some conclusions in Sec.~\ref{sec:conclusions}.

\section{Skyrme Lagrangian, Skyrme units, and choice of parameters}
\label{sec:skyrme}
\subsection{The model}
The ``standard'' Skyrme model is based on the non-linear sigma model, 
extended by a quartic interaction term and a pion mass term.
The instanton-induced form \cite{AtiyahManton89,AtiyahManton93} 
of the Skyrme fields can only be used in the (chiral) limit 
of zero pion mass, where the model is defined by the Lagrange density 
\begin{equation}
{\cal L}=\frac{f^{2}_{\pi}}{4}Tr(\partial_{\mu}U\partial^{\mu}U^{\dagger})
+\frac{1}{32 e^2}
Tr[U^{\dagger}\partial_{\mu}U,U^{\dagger}\partial_{\nu}U]
[U^{\dagger}\partial^{\mu}U,U^{\dagger}\partial^{\nu}U],
\label{eq:lagrangian}
\end{equation}
where $U$ is a unitary two-by-two matrix-valued field satisfying the
boundary condition $U=1$ at infinity. As has been discussed many times
before, this model has a topologically conserved quantum current.
The charge of this current is
identified with baryon number $B$. 

If we rescale the units of time and length, $x\rightarrow x/(ef_\pi)$,
(the so-called Skyrme units),
the Lagrange density takes on the slightly more convenient form
\begin{equation}
{\cal L}=\frac{f_{\pi}}{e}
\left(\frac{1}{4}Tr(\partial_{\mu}U\partial^{\mu}U^{\dagger})
+\frac{1}{32}
Tr[U^{\dagger}\partial_{\mu}U,U^{\dagger}\partial_{\nu}U]
[U^{\dagger}\partial^{\mu}U,U^{\dagger}\partial^{\nu}U]\right),
\end{equation}
 where ${f_{\pi}}/{e}$ is the Skyrme unit of energy.%

Finally the Skyrme Lagrangian can easily be reformulated in terms of the
Sugawara variables (Lie-algebra valued currents) $L$,
\begin{equation}
L_\mu = U^{-1}\partial_\mu U = i\,l_\mu^a \tau_a,
\label{eq:defLmu}
\end{equation}
and we have
\begin{equation}
{\cal L}=
\frac{1}{2}l_\mu^a l^\mu_a
+\frac{1}{4}\left[(l_\mu^a l^{\mu a})^2-l_\mu^a l_\nu^al^{\mu b} l^{\nu b}
\right]
\end{equation}

\subsection{Instanton-induced Skyrme fields}

As discussed in \cite{AtiyahManton93} one can derive a Skyrme field 
from a Yang Mills instanton 
field (called YMI-induced in this paper) 
by integrating the time component of the gauge potential,
\begin{equation}
U(\vec{x}) = C {\cal S} \left\{P \exp\left[\int_{-\infty}^\infty -A_4(\vec{x},t) dt \right]\right\}
C^\dagger .
\label{eq:holonomy}
\end{equation}
Here ${\cal S}$ is a constant matrix, chosen such that  $U$ decays 
to $1$ at infinity, and $C$ describes an overall grooming. For the current
work, where we shall only consider $C$ near the identity, it is convenient
to parametrise
\begin{equation}
C = \exp(i\vec \tau \cdot \vec \theta). \label{eq:deftheta}
\end{equation}
For the Jackiw-Nohl-Rebbi (JNR) instanton of charge $k$ we have
\cite{JNR}
\begin{equation}
A_4(\vec{x},t) = \frac{i}{2} \frac{\vec{\nabla}\rho}{\rho}\cdot\vec{\tau},
\;\;\;\rho=\sum_{l=1}^{k+1} \frac{\lambda_l}{|x-X_l|},
\label{eq:JNR}
\end{equation}
and we should use ${\cal S}=-I$, to obtain a field of positive baryon number $k$.
 
In order to solve for the YMI-induced value of $U$ we convert the integral 
(\ref{eq:holonomy}) to the solution of a differential equation. First introduce
\begin{equation}
\tilde{U}(\vec{x},\tau) = C{\cal S} \left\{
P \exp\int_{-\infty}^\tau -A_4(\vec{x},t)dt\right\}C^\dagger
.
\end{equation}
This function satisfies the differential equation
\begin{equation}
\partial_\tau \tilde{U}(\vec{x},\tau)
= -\tilde{U}(\vec{x},\tau)A_4(\vec{x},\tau),
\end{equation}
with initial condition $U(\vec{x},-\infty)={\cal S}$. The function
$U(\vec{x})$ is obtained as the limit for $\tau \rightarrow\infty$ of 
$\tilde U$.
We can work directly  with the Sugawara variables $L_\mu$, 
Eq.~(\ref{eq:defLmu}).
These can be calculated as the large $\tau$ limit of a quantity
$\tilde L_\mu$, which satisfies the differential equation
\begin{equation}
\partial_\tau \tilde L_\mu(\vec{x},\tau) = 
[A_4(\vec{x},\tau),\tilde L_\mu(\vec{x},\tau)] 
- \partial_\mu A_4(\vec{x},\tau).
\label{eq:Ltmu}
\end{equation}
Here the boundary condition is $\tilde L_\mu(\vec{x},-\infty)=0$.
By differentiating the  differential equations (\ref{eq:Ltmu})
we can obtain expressions for  derivatives of $L_\mu$ with respect to the 
instanton parameters $\lambda_l$ and $X_l$, which will be needed later.

The field $U$ defined in Eq. (\ref{eq:holonomy}) 
does not have an explicit time dependence.
There is an
implicit dependence, due to a possible variation of the 
instanton parameters as well as the unitary matrix $C$ (parametrised
by the three angles $\vec\theta$) with  time. 
Let us denote the parameters $\{\lambda_l,X_l,\vec\theta\}$ 
collectively by $\xi$. We then have
\begin{equation}
L_0 = {\dot \xi}^\alpha U^\dagger \partial_{\xi^\alpha} U
\equiv
 {\dot \xi}^\alpha L_{,\alpha}.
\label{eq:Leff}
\end{equation}
If we substitute this in the Lagrangian we obtain the form
\begin{equation}
L = T - V,
\end{equation}
with
\begin{equation}
V = \half \sum_{i,a}l^a_i l^a_i + \mbox{$\frac{1}{4}$}
\left[\left(\sum_{i,a}l^a_i l^a_i\right)^2 -\sum_{ij}
\left(\sum_{a}l^a_i l^a_j\right)^2\right],
\end{equation}
and
\begin{eqnarray}
T &= & \half \dot \xi ^\alpha \dot B_{\alpha\beta}\xi ^\beta ,\\
B_{\alpha\beta} & = &
\sum_a l_{,\alpha}^al_{,\beta}^a
+2\left[\sum_a l_{,\alpha}^al_{,\beta}^a\sum_{i,b}l^b_i l^b_i
-\sum_I\left(\sum_{a}l^a_i l^a_{,\alpha}\sum_{b}l^b_i l^b_{,\beta}\right)
\right].
\end{eqnarray}

The Lagrangian is quadratic in the time-derivatives due to the special nature 
of the Skyrme model. This also means that we have broken the Lorentz invariance
of the original equations, so  that (\ref{eq:Leff}) can only be
used to describe adiabatic (small velocity) motion.

In the  YMI-induced Ansatz we have thus replaced  the general matrix
$U(x)$, with an infinite number of parameters, by a 
form  parametrised by $3+5(k+1)$ parameters,
$C U(x|\xi)C^\dagger$.

\subsection{Skyrme units and model parameters}

\begin{table}[htb]
\caption{Different parameter sets as used in the Skyrme model}
\label{tab:skparms}
\begin{center}
\begin{tabular}{lll}
set & $f_\pi$ & e \\
\hline
ANW & 64.5 MeV & 5.45 \\
ANW$'$ & 81.7 MeV & 6.95 \\
R & 90 MeV & 4 \\
\end{tabular}
\end{center}
\end{table}

We take
three reasonable parameter sets in order to be able to
compare the effect of changes in the parameters.
The standard values for zero pion mass are those fitted
by Adkins, Nappi and Witten \cite{ANW} to the $N$ and $\Delta$ masses, and
are given in table \ref{tab:skparms} under the label ``ANW''.
Since in the instanton-induced Skyrmion the moment of inertia
differs considerably from its exact value, we propose a different
set of parameters under the label ``ANW$'$'', that give the
exact $N$ and $\Delta$ masses for the instanton-induced fields. Finally
we use a third set of parameters with realistic $f_\pi$ and
$e$. Of course we now no longer reproduce the masses.

\begin{table}[htb]
\caption{Our Skyrme units and their values}
\label{tab:skunits}
\begin{center}
\begin{tabular}{lllll}
parameter/ units & SU & ANW & ANW' & R \\
\hline{}
[length] & $1/(f_\pi e)$ & 0.562 fm & 0.347 fm & 0.548 fm\\{}
[energy] & $f_\pi/e$ &11.8 MeV & 9.28 MeV & 22.5 MeV\\{}
$\hbar$ & $e^2$ & 29.7 & 48.3 & 16\\
\end{tabular}
\end{center}
\end{table}

Now it may be useful to recapitulate our units, which we have done
in table \ref{tab:skunits}. Note that $\hbar$ is not one in 
Skyrme units, but $c$ is.

\section{Harmonic expansions}\label{sec:harmonic}
In these notes we shall use a harmonic expansion around (quasi-)stable
solutions of the Skyrme model. Starting from the Lagrangian 
(\ref{eq:lagrangian})
we can use the instanton-induced fields to formulate a harmonic approximation
of the form (we use $\xi$ for the deviation of the coordinates from
their equilibrium value)
\begin{equation}
H_{\rm ho} = E_0+\half B_{\alpha\beta}\dot\xi_\alpha\dot\xi_\beta
+ \half V_{,\alpha\beta} \xi^\alpha\xi^\beta.
\end{equation}
We can
diagonalise the harmonic Hamiltonian by making a linear transformation
to new coordinates
(see e.g. Ref. \cite{Goldstein})
\begin{eqnarray}
q^\mu &=& f^\mu_{,\alpha} \xi^\alpha ,\nonumber\\
\xi^\alpha &=& g^\alpha_{,\mu} q^\mu ,\nonumber\\
\end{eqnarray}
where $f$ and $g$ are the left and right eigenvectors of the matrix
\begin{equation}
M^\alpha_\beta = B^{\alpha\gamma}V_{,\gamma\beta},
\end{equation}
obeying the normalisation condition
\begin{equation}
g^\alpha_\mu f^\mu_\beta = \delta^\alpha_\beta,
\end{equation}
and with eigenvalues $(\hbar \omega_\mu)^2$.
The Hamiltonian can then be cast in the diagonal form
\begin{equation}
H = E_0 + \sum_\mu{}' \hbar\omega_\mu n_\mu + \sum_i \lambda_i^{-1} p_i^2,
\end{equation}
There the first sum runs over all non-zero eigenvalues, and the second
one only over the zero modes. Of course the quantised form of this same
Hamiltonian can easily be obtained by replacing $n_\mu$ by $n_\mu+1/2$,
interpreting $n_\mu$ and $p_i$ as operators. As is well known, see
e.g.\ Ref.~\cite{RingSchuck}, there is a correction to $E_0$ (due to 
normal ordering of the excitation operators and the non-zero expectation
value of the operators $p_i$ in the ground state), that usually
more than cancels the harmonic zero-point energy. Unfortunately,
these contributions are not easily evaluated in a field theoretical
context. See, however, Ref.~\cite{Moussalam}.

Using similar techniques, one can easily obtain expressions for the ground
state expectation values of observables. Of more direct interest when
interpreting the harmonic modes is the {\em change} in observables
related to a given mode. 
We shall only discuss quantities independent of the velocities. 
We expand
\begin{equation}
O = O_0 + \xi^\alpha O_{,\alpha} + \half \xi^\alpha\xi^\beta O_{,\alpha;\beta},
\end{equation}
$O_{,\alpha}$ etc. are nothing more than (covariant) 
derivatives w.r.t. $\xi^\alpha$. Of
course we can apply the chain rule to convert these to derivatives
w.r.t. $q^\mu$. To linear order one finds, for small $\delta\xi$,
\begin{equation}
\delta O = \delta \xi^\alpha O{,_\alpha} = \delta q^\mu g_{,\mu}^{\alpha}
O_{,\alpha}.
\end{equation}
From which we see that
\begin{equation} 
O_{,\mu} = g_{,\mu}^{\alpha} O_{,\alpha}
\end{equation}
is a good definition of the change of $O$ associated with the mode 
$\mu$. This expression is very useful in interpretation of the harmonic
 modes.
We can use it to define the change in baryon number, which is highly
enlightening.

In order to calculate the expectation value of the operator $O$ in
the ground state we now quantise the coordinate $\xi$ (or rather $q$) using
the standard harmonic oscillator rules
\begin{eqnarray}
q^\mu &=& \left(\frac{\bar{b}^\mu}{\bar{v}^\mu}\right)^{1/4}
\frac{(a^\dagger_\mu+a_\mu)}{\sqrt{2}},
\nonumber\\
p_\mu &=& -i\left(\frac{\bar{v}^\mu}{\bar{b}^\mu}\right)^{1/4}
\frac{(a^\dagger_\mu-a_\mu)}{\sqrt{2}},
\end{eqnarray}
where we have used the fact that $B_{\alpha\beta}$ and $V_{\alpha\beta}$
can be diagonalised simultaneously (with eigenvalues $\bar b$ and $\bar v$,
respectively).  We have $\hbar\omega_\mu=(\bar{b}^\mu\bar{v}^\mu)^{1/2}$.

The quantised form of the observable $O$ now becomes
\begin{equation}
\hat O = O_0 + \frac{1}{4} \sum_\mu{}' 
\left(\frac{\bar{b}^\mu}{\bar{v}^\mu}\right)^{1/2}
f^\alpha_\mu f^\beta_\mu O_{,\alpha;\beta}
+ {\rm normal~ordered~terms}.
\end{equation}
The second term can now interpreted as the zero-point-motion contribution
to $O$. Notice that everything above was done in the {\em intrinsic} 
frame (also called the body-fixed frame). In order to obtain the result
for an observable in the lab frame it still has to be averaged over
with a suitable rotational wave function \cite{RingSchuck}.

\section{Properties of the $B=1$ hedgehog}\label{sec:B=1}
The $B=1$ collective Hamiltonian is given by zero-modes only,
\begin{equation}
H = E_0 + \frac{\hbar^2}{2\Lambda}({I^2+S^2}) + \frac{1}{2 M} P^2.
\end{equation}
The parameters in $H$ are given in table \ref{tableB1}, where we both
show the results for the YMI-induced approximation, and for the numerical
solution of the full problem.

\begin{table}[htb]
\caption{Properties of the $B=1$ system.} \label{tableB1}
\begin{center}
\begin{tabular}{l|ll}
 & instanton-induced & exact \\
\hline
$E_0$ & 73.6 $f_\pi/e$ &  73.0 $f_\pi/e$ \\
$\hbar^2/\Lambda$ & 1/140.1 $f_\pi e^3$ & 1/106.6 $f_\pi e^3$\\
\end{tabular}
\end{center}
\end{table}

\section{Quantisation of the $B=2$ system.}\label{sec:B=2}
In this section we shall use the results obtained in Ref.~\cite{SkLACM}
for the $B=2$ system to make a set of approximations to the energy
of the deuteron of various levels of sophistication. We shall
start with a naive harmonic analysis, and end with an approximation
that goes beyond the calculation by Leese {\em et al}
\cite{LeeseMantonSchroers}.

\subsection{Harmonic analysis.}

In this section we shall discuss the harmonic analysis around the
doughnut in what has been called ${\cal M}_{12}$ by Manton et al.
This is a manifold that in the asymptotic regime has just
enough collective coordinates ($12 =6\times 2$)
to be able to project the individual Skyrmions onto nucleons. Three of
those describe the centre-of-mass motion, and do not couple. The 
remaining 9 modes describe the intrinsic dynamics.

The issue is the understanding of bound states in this channel, where
we restrict ourselves to the harmonic approximation. 
Since the doughnut has a continuous symmetry 
(generated by $M^{\rm int}_3=2I^{\rm int}_3+J^{\rm int}_3$)
we typically find two-dimensional pairs of modes when the quantum
number $m^{\rm int}_3$ is changed, and 1D when it remains zero.
For modes breaking the symmetry we find a centrifugal term of
the form $\hbar^2/cx^2 (M_3^{\rm int})^2$ (where $x$ is a coordinate describing
the deviation from equilibrium). This leads to a 2D harmonic
oscillator type spectrum, which can be given in the following form
\begin{eqnarray}
E &=& E_0 + \sum_{2D}\hbar\omega_i(n_i+m^{\rm int}_3+1) + 
\sum_{1D}\hbar\omega_i(n_i+1/2)\nonumber\\&&
+\frac{\hbar^2}{2{\cal I}_I}(I^2-(I^{\rm int}_3)^2)
+\frac{\hbar^2}{2{\cal I}_J}(J^2-(J^{\rm int}_3)^2)
+\frac{\hbar^2}{2{\cal I}_C}(I^{\rm int}_3-2J^{\rm int}_3)^2.
\end{eqnarray}
Let me first discuss
the classification of the modes:

\begin{table}[htb]
\caption{Harmonic frequencies around the $B=2$ solution in the YMI-induced
 form.}
\begin{center}
\begin{tabular}{llll}
Classification & deg. & $|\Delta m_3|$ & $\hbar\omega~/(f_\pi e)$\\
\hline
translational zero modes $(x-y)$& 2 & 1 & 0\\
rotational zero modes ($x-y)$& 2 & 1 & 0 \\
isorotational zero modes $(x-y)$ & 2 & 1 & 0 \\
translational zero mode $(z)$& 1 & 0 & 0 \\
mixed iso/rotational zero mode $(z)$& 1 & 0 & 0 \\
quadrupole mode & 2 & 2 & 0.352 \\
dipole mode & 2 & 1 &  0.419 \\
breathing mode & 1 & 0 & 0.524 \\
highest modes & 2 & 1 (?) & 0.756 
\end{tabular}
\end{center}
\end{table}

\begin{table}[htb]
\caption{Some properties of the $B=2$ solution in the YMI-induced form.}
\begin{center}
\begin{tabular}{ll}
$E_0$ & 141.18 $f_\pi/e$ \\
$\hbar^2/{\cal I}_I$&$1/114.2~f_\pi e^3 $ \\
$\hbar^2/{\cal I}_J$&$1/193.6~f_\pi e^3 $ \\
$\hbar^2/{\cal I}_C$&$1/65.2~f_\pi e^3 $ \\
\end{tabular}
\end{center}
\end{table}

We can use these number when estimating the energy in $M_{12}$ relative
to 2 $B=1$ Skyrmions. The ground state has intrinsic quantum numbers
$J=1,I=0,JJ^{\rm int}_3=0,I^{\rm int}_3=0$. We find the following
energy balance:
\begin{equation}
\begin{array}{rrclclcllc}
& && E_0 &&\half\sum\hbar\omega && E_{\rm rot} \\
&E_{B=2}  &=& 141.2 f_\pi /e &+&0.771 f_\pi e &+&\frac{1}{397.2} f_\pi e^3 \\
-&2E_{B=1} &=& 147.2 f_\pi/e&+& 0 &+&\frac{3}{2}\frac{1}{140.1} f_\pi e^3 &&
\raisebox{-1ex}{--} \\
\cline{1-8}
&\Delta E &=& -6.0 f_\pi/e&+&0.771 f_\pi e &+&-.00819~f_\pi e^3 
\end{array}
\end{equation}
For our three sets of parameters this takes on the values
117.7, 117.0 and 88.2 MeV. 

These estimates are all unbound, but
let us analyse the situation a little bit. We know first of all that
the moments of inertia are usually over-estimated by the YMI-induced Skyrme-fields.
Suppose we reduce all moments of inertia by a factor $106/140$, corresponding
to the ratio of the exact over the YMI-induced value. The ground-state energy
would then decrease by 27, 57 or 15 MeV (the middle value should be
disregarded, since this parameter set  was already corrected for this effect).

Furthermore we have ignored the fact
that we know that the behaviour in the attractive channel
direction (lowest 2 modes) is a lot less harmonic than in the other direction. 
The difference is due to the fact that in the attractive manifold the 
doughnut comes apart into two hedgehogs, and the doughnut is not strongly
bound. The depth of the well is 6 S.U. which takes on the values
71, 55.7 and  135 MeV for each of our three parameter sets; 
the harmonic energy quantum is 0.352$e^2$ S.U (the $e^2$ is really $\hbar$
in Skyrme units), which takes values 124, 200 and 127 MeV.
In the other direction we expect to reach states with 
energies of the order of the repulsive channel (maybe up to 1 GeV),
and the harmonic approximation should be good.

\begin{figure}[htb]
\epsfysize=7cm
\centerline{\epsffile{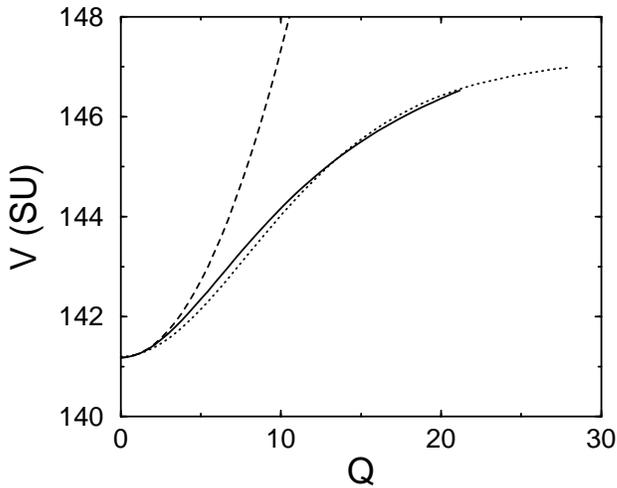}}
\caption{The potential energy as calculated in the attractive channel 
(solid line). The dashed line shows the harmonic approximation to this 
potential, and the dotted line a P\"oschl-Teller fit}
\label{fig:att_V}
\end{figure}

In order to see what might be the possible effect of the limited
well-depth we also calculate the binding energy of
the lowest state by approximating the potential by a P\"oschl-Teller 
form \cite{PT}.
This has the advantage that the problem is exactly soluble, and this
potential has only two parameters, which can be chosen to be
the depth and the harmonic frequency. If more information is known
one of the generalised potentials discussed by Ginocchio \cite{Ginocchio}
might be of some use. For the time being we use the form
\begin{equation}
V(x) = -\nu(\nu+1)/\cosh^2(\alpha x).
\end{equation}
Here we have chosen our coordinates such that $\hbar^2/m=1$. The parameters
$\nu$ and $\alpha$ can then be connected to the harmonic frequency and the
depth of the potential
\begin{eqnarray}
\nu&=& \frac{1}{2}(-1+\sqrt{1+8V_0/\alpha^2}),\nonumber\\
\alpha&=& \hbar\omega/\sqrt{V_0}.
\end{eqnarray}
For this exactly solvable potential we then find that the ground state occurs
at a zero-point energy
\begin{equation}
E_0 = \frac{\hbar\omega}{2}\left[\frac{\hbar\omega}{4V_0}
\left(\sqrt{1+16\left(V_0/\hbar\omega\right)^2}-1\right)\right]
\end{equation}
above the minimum of the potential. It is easy to see that this reduces
to the harmonic oscillator value in the limit $V_0\gg\hbar\omega$.

The reduction in the contribution for our lowest two modes would be
0.655, 0.518 and 0.768 for our three parameter sets. This corresponds to
43, 96 and 30 MeV reduction in binding. Taking all corrections 
together (we do not correct the middle value for the wrong value of 
the moment of inertia) we find 47, 21 and 43 MeV. These values
are reasonably close to zero, which is about the
best one could hope for such a crude calculation. We have summarised
our calculations in table \ref{tab:harmrel}.

\begin{table}[htb]
\caption{Summary of harmonic and related approximations for the deuteron energy
relative to the $B=2$ threshold, as discussed in the text.}
\label{tab:harmrel}
\begin{center}
\begin{tabular}{l|lll}
&ANW & ANW$'$ &R \\
\hline
harmonic & 118 MeV & 117  MeV& 88  MeV\\
corrected moment of inertia & 98 MeV & 117  MeV& 73  MeV\\
include anharmonicity & 47  MeV& 21  MeV& 43  MeV
\end{tabular}
\end{center}
\end{table}

\subsection{Quantisation in $M_{7}$ and extensions}
Leese, Manton and Schroers have constructed a Hamiltonian in the attractive
channel, where one has $7$ degrees of freedom. 
A quantisation of this Hamiltonian in the intrinsic frame
led to a binding energy
of $6$ MeV. From our calculation a similar Hamiltonian is available, and
it would be nice to independently check these predictions, and we have all
the necessary ingredients.\footnote{Note that there is an error in the 
collective coordinate $Q$ as presented in Ref.~\cite{SkLACM}, 
as well as a factor of two difference in the inertial parameters. 
The results presented here use the corrected $Q$.} 
We normalise our collective coordinate such that the mass
is $1$ in Skyrme units, and we then have to quantise the 
intrinsic collective Hamiltonian (see also Ref.~\cite{LeeseMantonSchroers})
\begin{eqnarray}
H  &=& \half\left(P^2+
\frac{(J^{\rm int}_x)^2}{a(Q)^2} + 
\frac{(J^{\rm int}_y)^2}{b(Q)^2} +
\frac{\left[J^{\rm int}_z-w(Q)I^{\rm int}_z\right]^2}{c(Q)^2} 
\right. \nonumber\\&&\left.
+\frac{(I^{\rm int}_x)^2}{A(Q)^2} + 
\frac{(I^{\rm int}_y)^2}{B(Q)^2} +
\frac{(I^{\rm int}_z)^2}{C(Q)^2}\right)
+ V(Q).
\end{eqnarray}
Using the standard geometrical quantisation, and the intrinsic quantum 
numbers compatible with the deuteron, $I=0, J=1, J_z^{\rm int}=0$, 
we obtain a ``radial'' equation of the form
\begin{equation}
\left[-
\half \frac{\hbar^2}{g^{1/2}}\partial_Q \left(g^{1/2} \partial_Q\right) +
\hbar^2\left(\frac{1}{2a^2}+\frac{1}{2b^2}\right) + V(Q)\right]
\psi(Q) = E \psi(Q).
\end{equation}
Here $g^{1/2}$ is the square-root of the determinant of the metric 
(mass matrix),
\begin{equation}
g^{1/2} = a b c A B C.
\end{equation}
\begin{figure}[htb]
\epsfysize=7cm
\centerline{\epsffile{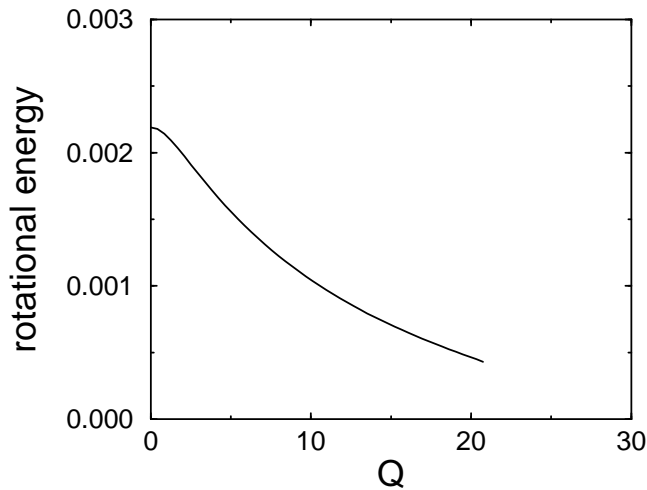}}
\caption{The contribution of the rotational kinetic energy to
the deuteron bound-state problem.}
\label{fig:att_rot}
\end{figure}
\begin{figure}[htb]
\epsfysize=7cm
\centerline{\epsffile{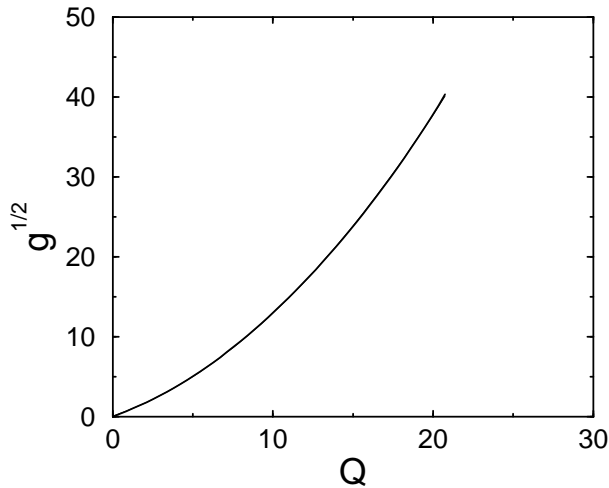}}
\caption{The metric factor $g^{1/2}$ in the kinetic energy, 
rescaled by a constant. }
\label{fig:att_g}
\end{figure}
The contribution of the rotational kinetic energy, which has units
$f_\pi e^3$, is plotted in Fig.~\ref{fig:att_rot}. We have subtracted
the asymptotic behaviour -- which corresponds to the system separating
into two hedgehogs locked in the attractive configuration. 
This is not the same as subtracting the kinetic energy of two hedgehogs
in nucleon states. This would require that one includes the harmonic motion
changing to zero-mode motion as the hedgehogs pull apart. 
This can (and should) only be done for scattering states \cite{WaletScatt}.
For a bound-state calculations  this
part is missing (see below), one should only subtract so much rotational
zero-point energy as to put the ionisation barrier at 0.

The metric factor
$g^{1/2}$ (scaled by a constant) is plotted in Fig.~\ref{fig:att_g}. It is
very well fitted by a form $\alpha Q + \beta Q^2$. 
The proportionality to $Q$ for small $Q$
agrees with the two-dimensional harmonic nature of the motion there. The
asymptotic behaviour agrees with the result by Leese {\em et al}. 
Mathematically
one can interpret this as a system that changes it nature from two-dimensional
at small $Q$ to 3D at large separations. We have fitted all the relevant
quantities as a function of separation, taking into account the exact
behaviour for large separation. 
This allows one to solve
the Schr\"odinger equation numerically. 
Some selected results are presented in 
table \ref{tab:numsol}.

\begin{table}[htb]
\caption{The energy of the lowest bound state in units of $f_\pi/e$
as a function of $e$. The three columns denote the pure potential (no
rotational energy), potential plus rotational energy and the additional effect
 of also adding the harmonic zero-point energies for the other modes in 
$M_{12}$. A dash indicates no solution.\label{tab:numsol}}
\begin{center}
\begin{tabular}{llll}
$e$ &       potential &        rotational &        all\\
\hline
1 &      -5.67       &    -5.67    &       -5.49\\
2 &      -4.94       &    -4.63    &       -4.43\\
3 &      -3.95       &    -3.83    &       -2.47\\
4 &      -2.67       &    -2.32    &       -0.30\\
5 &      -1.28       &    -0.65    &       ---\\
6 &      -.0044      &    ---      &       ---\\
\end{tabular}
\end{center}
\end{table}

\begin{figure}[htb]
\epsfysize=7cm
\centerline{\epsffile{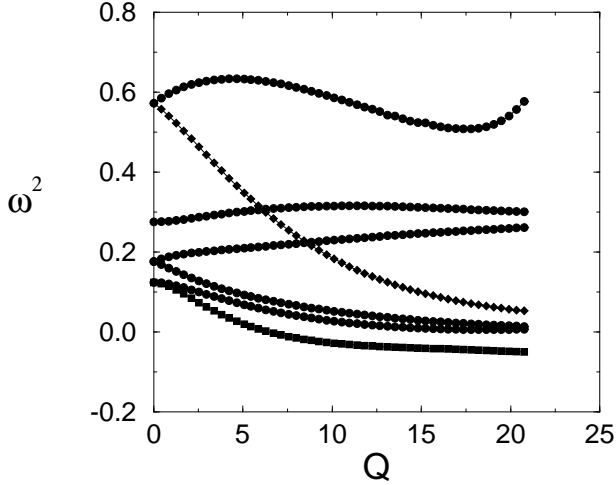}}
\caption{A correlation diagram, showing how the local harmonic frequencies 
are correlated as we move from the doughnut in the attractive channel.}
\label{fig:att_correl}
\end{figure}

\begin{figure}[htb]
\epsfysize=7cm
\centerline{\epsffile{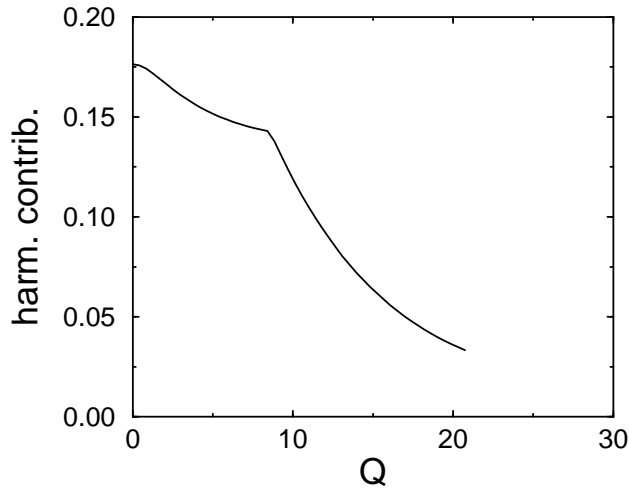}}
\caption{The harmonic energy as added to the collective Hamiltonian.}
\label{fig:Eharml}
\end{figure}

In order to mimic the effects of quantising in the full 12-dimensional 
configuration space corresponding to full translational and iso-rotational
zero modes of the two hedgehogs that appear at large separation we add 
the harmonic zero-point energies for these modes. Let us first study
a correlation diagram, that shows all the non-zero modes as they develop
from the doughnut, see Fig.~\ref{fig:att_correl}. As one can see over there
counting modes causes some problems.  The lowest two modes correspond to 
the attractive channel, and the next two modes are the ones describing the
two remaining fluctuations in $M_{12}$. The problem lies in the level crossing
at $Q=8.8$. If this were an avoided crossing, we would happily follow the 
lowest mode. Unfortunately it is not: there is quite a lot of symmetry
in the attractive channel and the two modes appear to have {\em
different}
symmetries. This may be an artifact of the YMI-induced Ansatz, but it seems more likely
that this is a real effect. Since this symmetry is closely
linked to the attractive channel, it is highly probable that if we 
deform the system
a little bit away from the attractive channel, we get an avoided crossing.
The crossing point would then correspond to what has been termed a diabolic
point in the theory of Berry's phases, and lead to a monopole type geometric 
phase. For the time being we ignore this effect and shall sum the zero-point 
energy of the two lowest modes, as shown in Fig.~\ref{fig:Eharml}.

Results including this harmonic contribution are give in the
third column of table \ref{tab:numsol}. Taking all the results together,
we are forced to conclude that is almost impossible to find a bound
state for a reasonable set of parameters when disregarding the harmonic
motion, whereas when including the harmonic motion a value for $e$ of at
most four is allowed. This is a contradiction of the claim of $6~\rm MeV$
binding energy by Leese {\em et al} \cite{LeeseMantonSchroers}. They used
a set of Skyrme parameters appropriate to a finite pion mass, however,
without examining the parameter dependence. The fact that there is a bound
state close to zero energy, or a resonance slightly above is the most
important issue. One cannot consider the model to be right on scales
of the order of a few MeV out of 2GeV, especially not without including
loops and the related counterterms \cite{Moussalam}.

\section{The structure of the $B=3$ collective manifold}\label{sec:LACMB=3}

\subsection{Harmonic modes around the tetrahedron}
For the $B=3$ case we use the YMI-induced+JNR form as used by Leese and Manton
\cite{LeeseManton}. This is not based on the most general instanton field 
but is probably sufficiently general for our purposes.

The instanton field is generated by four poles, arranged in an tetrahedron.
All weights are equal. We choose 
\begin{eqnarray}
(\lambda_1,X_1) &=& (1/4,-R,-R,-R,0),\nonumber\\
(\lambda_2,X_2) &=& (1/4,-R,R,R,0),\nonumber\\
(\lambda_3,X_3) &=& (1/4,R,-R,R,0),\nonumber\\
(\lambda_4,X_4) &=& (1/4,R,R,-R,0),
\end{eqnarray}
where $R$ is a parameter determining the size of the tetrahedron.

\begin{figure}
\epsfxsize=8cm
\centerline{\epsffile{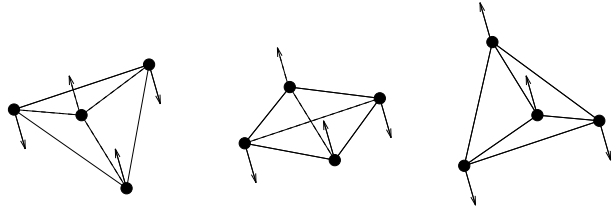}}
\caption{Tunnelling from one tetrahedron to another.}\label{fig:tunnel}
\end{figure}

Since all continuous symmetries are broken, we find 9 zero modes. The remaining
modes can be classified as follows (The interpretation and classification
of these modes is discussed in more detail in the appendix):
\begin{itemize}
\item[10-11]
$\hbar\omega=0.318f_\pi e$.\\
These two modes correspond to the motion towards the $B=3$ doughnut, 
See Fig.~\ref{fig:tunnel} for a sketch.

\item[12-14] 
$\hbar\omega=0.332f_\pi e$.\\
These modes are the beginning of a trajectory towards the pretzel
mentioned in the introduction.

\item[15-17]
$\hbar\omega=0.469f_\pi e$.\\
It is highly probable that these modes describe the separation in 
a $B=2$ doughnut and a single hedgehog.

\item[18] 
$\hbar\omega=0.483f_\pi e$.\\
This is the breathing mode.

\item[19-21]
$\hbar\omega=0.813f_\pi e$.
These are the only modes that mix with iso-spin rotations. They describe
a complex high-energy excitation.
\end{itemize}

If we assume that we need 18 modes to describe the $B=3$ collective 
coordinates, we notice that there is a clear gap in the spectrum just at that
critical point. We have not used the most general instantons here, and it is 
unlikely but not impossible that this situation might change if we switch
to the AHDM instanton \cite{AHDM}.

\begin{table}[htb]
\caption{Some properties of the $B=3$ solution in the YMI-induced form.\label{tabB3}}
\begin{center}
\begin{tabular}{ll}
$E_0$ & 206.6 $f_\pi/e$ \\
$\hbar^2/{\cal I}_I$&$(113)^{-1}~f_\pi e^3 $ \\
$\hbar^2/{\cal I}_J$&$(357)^{-1}~f_\pi e^3 $ \\
$\hbar^2/{\cal I}_C$&$(-633)^{-1}~f_\pi e^3 $ \\
\end{tabular}
\end{center}
\end{table}

As said before, 
the lowest non-zero modes of the tetrahedron correspond to  motion towards 
the $B=3$ doughnut, where the line connecting any of the corners
of the tetrahedron moves closer to the line connecting the
other two corners. When these lines cross we have reached the $B=3$ doughnut 
which in our Ansatz is described by four poles arranged in a square.
Actually one would expect three such modes, one for each coordinate axis. 
However, the sum of these modes is the breathing mode, which is in a different
representation of the tetrahedral group, and thus we find only
two modes (see also our model discussion below). 
Following these modes naturally leads one 
to study the doughnut as well. It sits on 
the top of a saddle-point in energy. It has two unstable modes, whereas 
we expected only one corresponding to the path towards the tetrahedron. 
Indeed this is the most unstable mode. The other unstable mode leads to 
yet another saddle point. This solution has the double-hole structure of 
a pretzel. The potential energy of all these states, 
{\em i.e.}, ignoring contributions from the rotational kinetic energy, 
is not very different
\begin{eqnarray}
&&E_{\rm tetrahedron} =206.56 f_\pi/e  ,\;\;
E_{\rm pretzel} = 211.54 f_\pi/e   ,\;\;
E_{\rm doughnut} =214.06 f_\pi/e ,\nonumber\\
&&3E_{B=1}=220.8 f_\pi/e, E_{B=2}+E_{B=1}=214.8 f_\pi/e.
\end{eqnarray}

As one can see from table \ref{tab:sizes} the sizes of the different 
configurations are different. Here we have given the values of the
integrals 
\begin{equation}
\langle r_i^2 \rangle = \int r_i^2 B(\vec r) d^3r,
\end{equation}
which is the standard definition for these quantities. 

\begin{table}[htb]
\caption{Sizes of the different shapes for $B=3$ in S.U.}\label{tab:sizes}
\begin{center}
\begin{tabular}{lllll}
solution & $\langle xx\rangle$ & $\langle yy\rangle$ & $\langle zz\rangle$ & $\langle r^2\rangle$\\
\hline
tetrahedron & 3.043 & 3.043 & 3.043 & 9.129 \\
pretzel & 9.225 & 2.436 & .916 &   12.577\\
doughnut & 4.974 & 4.974 & .878 & 10.836\\
\end{tabular}
\end{center}
\end{table}

\begin{figure}[htb]
\epsfysize=6cm
\centerline{\epsffile{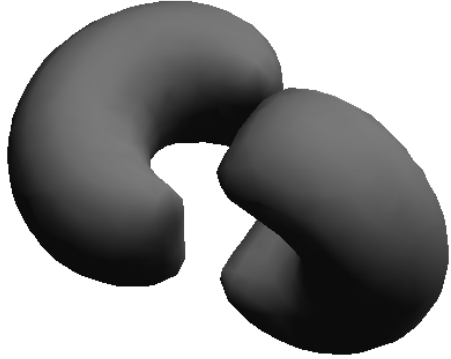}}
\caption{An example of the baryon density found when following
the path through the doughnut and the tetrahedron, to a point
beyond the tetrahedron}
\label{fig:twist}
\end{figure}

Let us first consider the paths connecting the different solutions.
As we stated there is a path connecting the two independent 
tetrahedral configurations through a saddle point that looks like a doughnut.
Actually there are three such paths, corresponding to the three
``independent'' doughnuts with rotational symmetry around each of the
three coordinate axes.
Such a path is very similar  to the geodesic with twisted symmetry
found in the scattering of three collinear monopoles by Houghton and
Sutcliffe \cite{HouSut}. It corresponds to motion where the Skyrmion
breaks apart into three $B=1$ hedgehogs in an attractive (and therefore
twisted) configuration. The only thing is that in our case the Skyrmion
does not go into into this form, but rather breaks up into two $B=3/2$
``bananas'' as shown in Fig.~\ref{fig:twist}. Since solutions with fractional
baryon number do not exist in the Skyrme model -- they have infinite energy --
this leads to
an infinite energy in the limit of infinite separation. It is plausible
that the use of the JNR instanton, instead of the full AHDM data
is responsible for this effect -- It is hard to imagine how to
describe the desired configuration in terms of the JNR data. For the motion
{\em in between}
 the doughnut and hedgehog the JNR form appears to be reasonable.

\begin{figure}[htb]
\epsfysize=7cm
\centerline{\epsffile{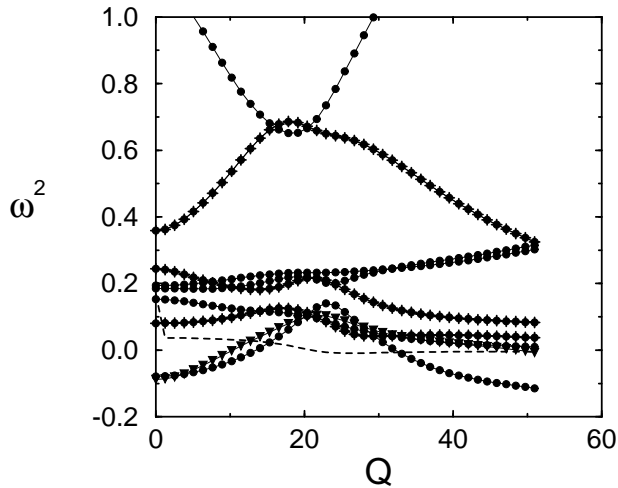}}
\caption{A correlation diagram, showing how the local harmonic frequencies 
are correlated as we move from the doughnut to the tetrahedron 
(at $Q\approx 20$) and beyond. The lines with pluses and circles are
doubly degenerate. The triangles are used to label the mode
actually followed. The dashed line shows the new (zero-)mode appearing
when we move away from the doughnut. Note that this connects to a non-zero mode
at the doughnut. This is probably another deficiency of the JNR 
parametrisation.}
\label{fig:corel_dn}
\end{figure}

\begin{figure}[htb]
\epsfysize=7cm
\centerline{\epsffile{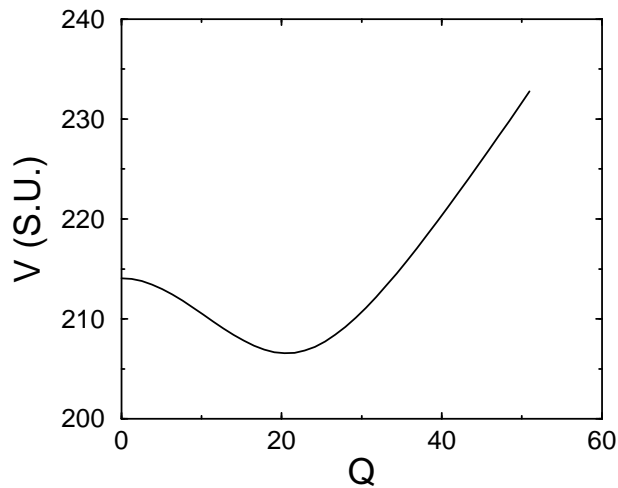}}
\caption{The potential energy as a function of the coordinate $Q$,
as we move from the doughnut to the tetrahedron  
(at $Q\approx 20$) and beyond. Note the steep rise of the potential, above
the ``ionisation'' limit of 220.8 S.U., when the system splits into two
$B=3/2$ ``bananas''.}
\label{fig:E_dn}
\end{figure}

In Fig.~\ref{fig:corel_dn} we show how the local harmonic frequencies 
change as we move from the doughnut to the tetrahedron and beyond.
As usual we have fixed the normalisation of the coordinate $Q$ by
requiring that the mass (which we usually call $\bar B$) equals one.
There are no major exciting structures along the path, and one can see
that the mode we followed (denoted by the triangles) interpolates smoothly
between the tetrahedron and doughnut. Furthermore we find that several
modes are doubly degenerate for all $Q$ due to the  symmetry
in the configurations along the path. In this case ignoring the
weights in the JNR instanton is no limitation; they are constant and
all equal for all points along the path. The potential energy along the
path is given in Fig.~\ref{fig:E_dn}.

Once one has the doughnut, one can study its unstable modes. It has two,
the one discussed above, and one leading to the pretzel. 
This latter solution is described by a rhombic arrangement
of the poles, so that for each doughnut we have two neighbouring pretzels,
obtained by compressing any of the two diagonals of the square 
arrangements of poles describing the doughnut. In this
case the weights in the JNR potential are also no longer equal, but this
need not concern us here. The pretzel has one unstable mode, which 
connects to the tetrahedron, as we discovered through our mode following
approach. 

\begin{figure}[htb]
\epsfysize=7cm
\centerline{\epsffile{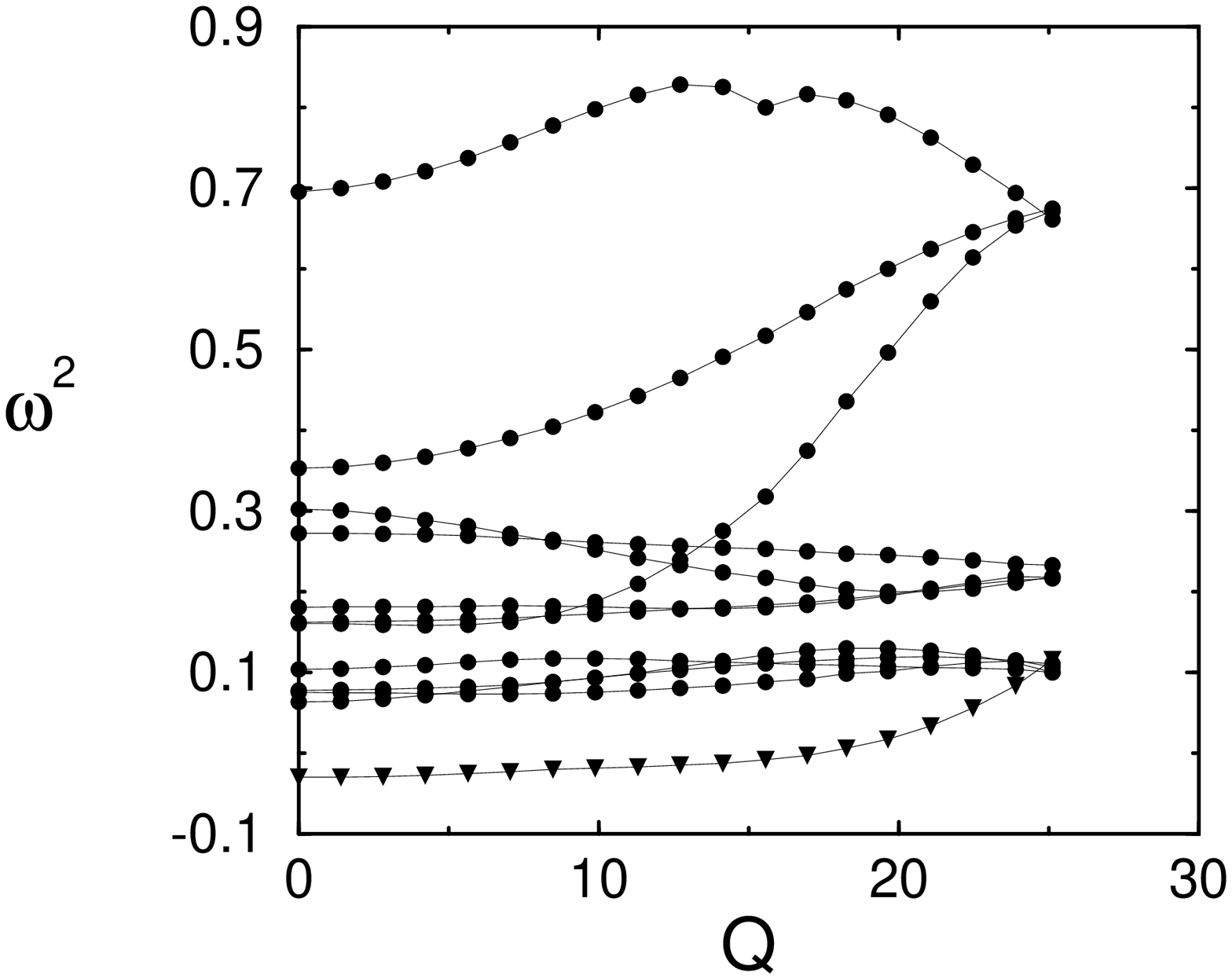}}
\caption{A correlation diagram, showing how the local harmonic frequencies 
are correlated as we move from the pretzel ($Q=0$) to the tetrahedron 
(at $Q\approx 25$).}
\label{fig:corel_pr}
\end{figure}

A very important property of the path between the tetrahedron and the
pretzel is that at all points the Skyrme fields have a trivial reflection
symmetry under coordinate reflections, namely $\vec \phi(R_i \vec x)
= R_i \vec\phi (\vec x)$.

\begin{figure}[htb]
\epsfysize=4cm
\centerline{\epsffile{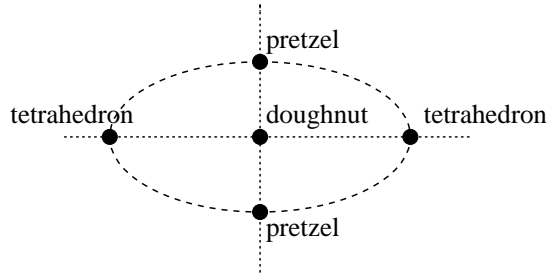}}
\caption{A schematic sketch of a few of the connections through a given
doughnut. For each doughnut in one of the three coordinate planes, there
are two associated pretzels, elongated along one of the two coordinate
axes in the plane, and connected by a valley. There is a connection
of the doughnut to two tetrahedra with different orientations. There
also exists a closed valley connecting the tetrahedra through the
pretzels.}
\label{fig:lands}
\end{figure}

In Fig.~\ref{fig:corel_pr} we show how the local harmonic frequencies 
change as we move from the pretzel to the tetrahedron.
Actually we find that at the tetrahedron the mode followed
(again denoted by squares) connects to the lowest triplet of
modes in the doughnut. Since each of the modes of the triplet
describes a circuit through two pretzels and another tetrahedron,
as denoted schematically in Fig.~\ref{fig:lands}.
Note that the lowest mode for the
tetrahedron is the one connecting to the doughnut, which is the higher
of the two saddle points. Obviously this figure only sketches one
out of three connections, the whole structure of the collective
manifold being highly complex.

\section{Quantisation of the collective coordinates}
\label{sec:quantB=3}
\subsection{Harmonic approximation}
As in the case of the $B=2$ system, the most straightforward way to
get a ``quantum'' result for $B=3$ is to quantise in harmonic
approximation. Near the tetrahedron the collective Hamiltonian leads
to an expression for the energy of the following form
\begin{eqnarray}
E &=& E_0 + 
\sum_{\rm modes}\hbar\omega_i(n_i+1/2)\nonumber\\&&
+\frac{\hbar^2}{2{\cal I}_I}I^2
+\frac{\hbar^2}{2{\cal I}_J}J^2
+\frac{\hbar^2}{2{\cal I}_C}\vec I \cdot \vec J.
\end{eqnarray}

The ground state has intrinsic quantum numbers
$I=1/2,J=1/2,K=0$. Replacing the inner product of $I$ and $J$ with
the square of the grand-spin operator $K$,
we find the following energy balance:
\begin{equation}
\begin{array}{rrclclcllc}
& && E_0 &&\half\sum\hbar\omega && E_{\rm rot} \\
&E_{B=3}  &=&  206.6f_\pi /e &+& 1.76 f_\pi e &+&.00353 f_\pi e^3 \\
-&3E_{B+1} &=& 220.8 f_\pi/e&+& 0 &+&\frac{9}{4}\frac{1}{140.1} f_\pi e^3 &&
\raisebox{-1ex}{--} \\
\cline{1-8}
&\Delta E &=& -14.2 f_\pi/e&+& 1.76 f_\pi e &-& 0.0125 f_\pi e^3 
\end{array}
\end{equation}
For our three sets of parameters this takes on the values $319.2$,
$386.5$ and $242.1$ MeV. Here we once again see the effect of the over-estimate
of quantum corrections. Notice that without these corrections the value
would be $-168$, $-132$ and $-320$ MeV, respectively. 
Since we expect anharmonicities to play a large r\^ole, 
we shall now study
the effect of large amplitude fluctuations.

\subsection{Approximate large-amplitude dynamics}
As a fast and relatively simple way
to study the modes one can assume that one can independently 
quantise the motion
along the straight line and the ellipse, since these
represent orthogonal modes at the tetrahedron.

We first look at the paths through the doughnut. We have constructed a simple
model, based on the VSEPR model \cite{VSEPR},
as has been used in the chemistry of molecules. It is based on putting
particles on a sphere, and letting them interact through a
pairwise repulsive interaction, which is taken to be  some power of the inverse
of the separation of the particles.
If we do that in our problem we should  constrain the JNR poles 
on a sphere. The minimal energy configuration will be a tetrahedron,
and there will be a saddle point at the a square configuration of poles,
corresponding to the doughnut.
The configuration space is an eight-dimensional manifold. 
Three of those are the rotational zero modes.
At the tetrahedron the remaining modes are a pair of degenerate modes 
describing the motion towards the squares and three further-non-zero modes. 
These last three are of no interest, but can be eliminated by imposing the 
additional constraint that
the distance between two pairs of points is equal, and that the line
connecting the two mid-points of each pair goes through the centre of 
the sphere.

\begin{figure}[htb]
\epsfysize=4cm
\centerline{\epsffile{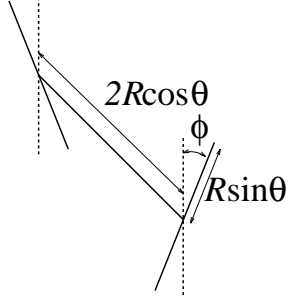}}
\caption{Coordinates used in our VPSER-based model. \label{fig:model1}}
\end{figure}
\begin{figure}[htb]
\epsfysize=6cm
\centerline{\epsffile{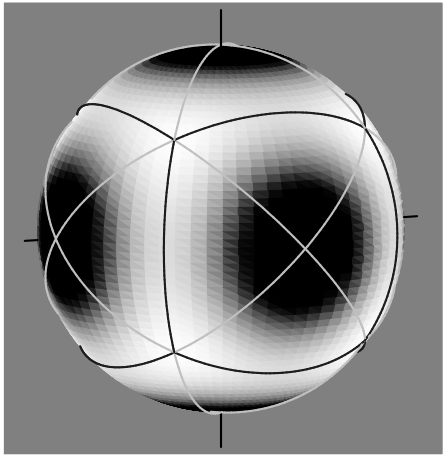}}
\caption{The potential energy (\protect{\ref{eq:pot}}) plotted
as a function of $\theta$ and $\phi$. The white areas denote lowest
potential energy, the black areas are where the potential diverges.
The black lines show the tunnelling paths between different minima;
the gray lines are the continuation of the black ones into 
the repulsive regime.
\label{fig:pot1}}
\end{figure}

With these constraints, using the coordinates specified in 
Fig.~\ref{fig:model1}, we obtain
\begin{eqnarray}
\vec{r}_1 &=& R D(\Omega) (\sin\theta\cos\phi,\sin\theta\sin\phi,\cos\theta), \\
\vec{r}_2 &=& R D(\Omega) (-\sin\theta\cos\phi,-\sin\theta\sin\phi,\cos\theta),\\
\vec{r}_3 &=& R D(\Omega) (\sin\theta\cos\phi,-\sin\theta\sin\phi,-\cos\theta),\\
\vec{r}_4 &=& R D(\Omega) (-\sin\theta\cos\phi,\sin\theta\sin\phi,-\cos\theta).
\end{eqnarray}
(Here $D(\Omega)$ is a rigid body-rotation corresponding to the three 
zero-modes). From this we can calculate the classical Lagrangian,
\begin{eqnarray}
{\cal L} &=& \half \sum_i \dot{\vec{r}}_i^2 - \sum_{i<j} 1/|r_i-r_j|^{2n}
\nonumber\\
&=& 2 R^2({\dot\Omega}^2 + {\dot \theta}^2 + \sin^2\theta{\dot \phi}^2)
\nonumber\\&&
-2/4^n\left(
1/(\sin^2\theta)^{n}+
1/(\cos^2\theta +\sin^2\theta\sin^2\phi)^{n}+
\right.\nonumber\\&&\left.
1/(\cos^2\theta +\sin^2\theta\cos^2\phi)^{n}
\right).
\label{eq:pot}
\end{eqnarray}
A lot of insight can be gained by plotting the potential as a function of
$\theta$ and $\phi$. The kinetic energy is proportional to ${\vec L}^2$,
which allows the interpretation of the configuration space as a sphere.
In order to plot the potential on this sphere, we have colour-coded the
values of $V$. In Fig.~\ref{fig:pot1} one can see this potential.
We have also indicated the tunnelling paths from one minimum to the next.
Here one can see why we have only two independent modes, and 
not three: one can not have three {\em orthogonal} paths in two dimensions.

\begin{figure}[htb]
\epsfysize=6cm
\centerline{\epsffile{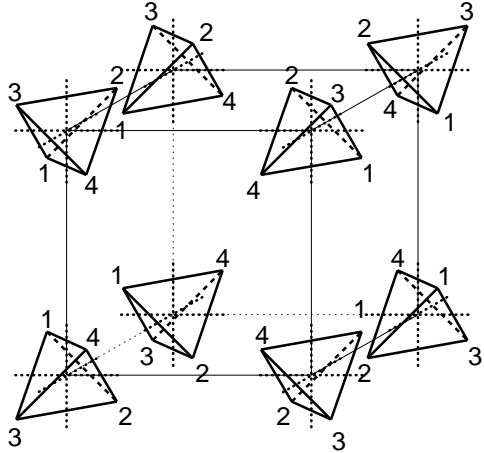}}
\caption{Orientation of the tetrahedra corresponding to each of the
vertices of the cube in Fig.~\protect{\ref{fig:cube}}\label{fig:quasi_rot}}
\end{figure}

One has to be somewhat careful how to interpret this sphere. In essence
there are only two independent orientations of the tetrahedron (the minimum),
which are related through an element of the octahedral group that is not
in the tetrahedral group. A sketch of that is given if 
Fig.~\ref{fig:quasi_rot}. As can be seen there, adjacent vertices have
independent tetrahedra, whereas all next-to-nearest neighbours correspond
to the same tetrahedron and a relabelling of the poles. 
This relabelling does nothing for the Skyrme
fields, and we are thus forced to identify these configurations as the same
point. This tetrahedral invariance means that the quantum mechanics
is actually restricted to a quarter of the sphere, with slightly complex 
boundary conditions. The potential energy has an even higher symmetry,
however. Its fundamental domain, 1/24th of the sphere, is mapped onto
the whole sphere by the $O_h$ group. 
Qualitatively one would expect that the lowest state has maximal symmetry.
 This is realized if we make a symmetric combination
of individual wave functions located in each minimum, leading to a
cubic arrangement of wave functions. Note that this is not directly related
to the pole positions; each minimum corresponds to a different tetrahedron.
They can be arranged in two groups of four. Within each group
the tetrahedra are related by a permutation of the poles (or an element
of the tetrahedral group). The two groups are related by a $90^\circ$
rotation, an element of the octahedral group. Since the pion field
transforms non-trivial under all these transformations, we end
up with a configuration explicitly symmetrised with respect to the
full octahedral group.

The tunnelling to the pretzel is somewhat harder to describe in the same
terms, but that may not be necessary. We find that there are three -- periodic
-- paths that describe tunnelling possibilities from one tetrahedron to
one that is related by a reflection in a plane through the centre
of the tetrahedron and parallel to two of the edges (an element of the
octahedral and not of the tetrahedral group).
Since these have a saddle point at lower energy than
the paths through the doughnuts, one expects that
they are even more important than those solutions. In essence these
paths can be visualised as the edges of a cube, where each closed path 
is equivalent to one edge.

\begin{figure}[htb]
\epsfysize=4cm
\centerline{\epsffile{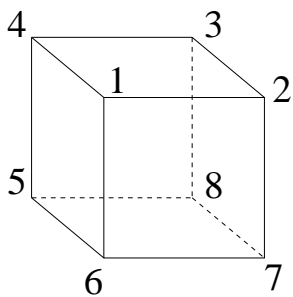}}
\caption{Labelling of the vertices of the cube. Each corresponds to a 
different realization of the tetrahedron
\label{fig:cube}}
\end{figure}
If we label the different edges of the cube as in Fig.~\ref{fig:cube}
we can estimate the spectrum using a simple H\"uckel model, where we only
include the nearest-neighbour interaction
diagonalising a simple 8-by-8 Hamiltonian matrix
\begin{equation}
H = \left(\begin{array}{llllllll}
-E_0 & -X & 0  & -X & 0 & -X & 0  & 0  \\
-X & -E_0 & -X & 0  & 0 & 0  & -X & 0 \\
0  & -X & -E_0 & -X & 0 & 0 &0 & -X  \\
-X & 0 & -X & -E_0 & -X & 0 & 0 & 0 \\
0 & 0 & 0 & -X & -E_0 & -X & 0 & -X \\
-X & 0 & 0 & 0 & -X & -E_0 & -X & 0 \\
0 & -X & 0 & 0 & 0 & -X & -E_0 & -X \\
0 & 0 & -X & 0 & -X & 0 & -X & -E_0
\end{array}\right) .
\end{equation}
where we assume all the eight configurations to be orthogonal.
$E_0$ is an  estimate for the binding energy within a single well,
and $X$ is a mixing matrix element (whose sign will be positive).
Since we  have to identify even and odd tetrahedra, the only
two eigenvectors of this matrix allowed are the symmetric and antisymmetric
ones. The eigenvalues are 
$-E_0\pm 3X$, and 
$-E_0 - 3X$ has eigenvector $(1,1,1,1,1,1,1,1)$.

\begin{figure}[htb]
\epsfysize=7cm
\centerline{\epsffile{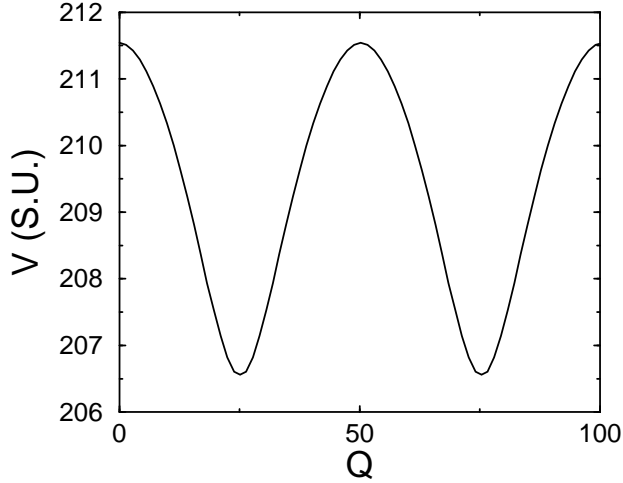}}
\caption{The potential energy for a full period of motion along the circuit
through the pretzels and the tetrahedron.}
\label{fig:pot_pr}
\end{figure}

\begin{figure}[htb]
\epsfysize=7cm
\centerline{\epsffile{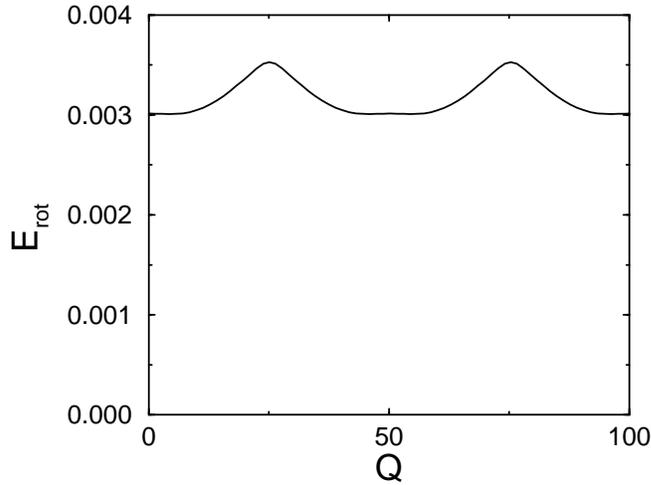}}
\caption{The expectation value of the
rotational energy  for a full period of motion along the circuit
through the pretzels and the tetrahedron.}
\label{fig:rot_pr}
\end{figure}

Since this concerns a lower barrier, it is of greatest
relevance to the large amplitude motion. The major issue here is that
the harmonic oscillator length parameter for each of the wells is comparable
to the length of the circuit for any reasonable value of Skyrme-model 
parameters. 
In this case the wave function in the circuit becomes approximately constant, 
and the energy is the average of the
potential along the circuit. Since the contribution of the rotational
kinetic energy tends to flatten the potential even more,
 Fig.~\ref{fig:rot_pr}, this only reinforces this behaviour.
If we just naively solve the one-dimensional problem on the circle,
we find an energy of $209.5 f_\pi/e+.00319 f_\pi e^3$, 
a considerable rise in energy.
Furthermore, since the pretzel is a lot bigger than the tetrahedron, we find
an enhanced value for the r.m.s. radius (by about $10\%$).

\begin{figure}[htb]
\epsfysize=5cm
\centerline{\epsffile{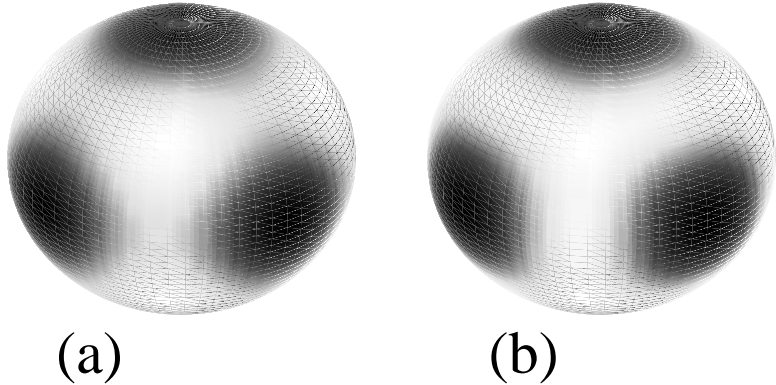}}
\caption{Two wave functions on the sphere. (a) $x^2y^2+y^2z^2+z^2x^2$
and (b) $x^4y^4+y^4z^4+z^4x^4+(x^2y^2+y^2z^2+z^2x^2)/2$
\label{fig:wfs}
}
\end{figure}

For the paths through the doughnut we cannot readily make such an estimate.
Even though the harmonic oscillator length parameter is again close to the
length of the path, which
would imply a similar estimate as above, we cannot ignore the two-dimensional
nature of the problem here. If we return to our VSEPR model, the simplest
symmetric wave function, that has zeros at the infinities of the potential is
$x^2y^2+y^2z^2+z^2x^2$ (it is easier to write the wave function in terms
of Cartesian tensors than as spherical harmonics). That wave function,
as plotted in Fig.~\ref{fig:wfs}a,
carries non-zero kinetic energy, and a straightforward calculation shows
that
\begin{equation}
\langle K \rangle = \frac{16}{5} \hbar^2.
\end{equation}
If we make the wave function more strongly centred along the tunnelling paths,
as in Fig.~\ref{fig:wfs}b, the expectation value of $K$ increases (
to $3.63\hbar^2$ for
this example). The best possible estimate for the contribution to the ground 
state energy from the two lowest modes is
\begin{equation}
\Delta E = \frac{1}{R^2}\langle K \rangle + \frac{1}{L}\int (V-E_0) dl
+\frac{1}{L}\int E_{\rm rot} dl.
\end{equation}
Here $R$ is the radius of the sphere associated with a tunnelling path of
length $L$, $L=2\arctan(\sqrt{\half}) R$. 
The integral is along one of the tunnelling paths. 
There are
many objections one can make to this approximation (the real collective
surface is not exactly a sphere, we ignore the zero-point motion orthogonal
to the path, etc.) but we have been unable to come up with a reasonable
alternative. We estimate that $\langle L^2 \rangle \approx 3.5 \hbar^2$,
and calculate the remaining quantities from our results. 

 Our final model is an independent quantisation 
of the five lowest non-zero modes, treating the remaining four in harmonic 
approximation. The potentials for the lowest five modes includes
the {\em difference} of the local harmonic energy and the harmonic energy
at the minimum,
\begin{equation}
\half \sum_{i=5}^9 \hbar\omega(Q)-\half \sum_{i=5}^9 \hbar\omega(0).
\end{equation}
The rotational energy is treated in a similar way, i.e., we subtract
the result at the minimum. The final result for the ground state energy
is
\begin{equation}
E_{gs} = E_{\rm tetrahedron}+\half \sum_{i=5}^9 \hbar\omega(0)+E_{\rm rot}(0)
+\Delta E^{(12)} + 3 \Delta E^{(3)}.
\end{equation}
Here we exhibit that we independent quantise modes three to five, and
have an explicit two dimension problem for the lowest two modes.
The various quantities in this expression are found to be 
(obtained in ways discussed above)
\begin{eqnarray}
E_0 &=& 206.6\frac{f_\pi}{e},\nonumber\\
\half \sum_{i=5}^9 \hbar\omega(0)&=& 0.941 f_\pi e, \nonumber\\
E_{\rm rot}(0)&=&0.00353 f_\pi e^3, \nonumber\\
\Delta E^{(12)} &=& 3.8 f_\pi/e -0.037 f_\pi e + 0.0028 f_\pi e^3, \nonumber\\
\Delta E^{(3)}&=& 2.9 f_\pi/e -0.041 f_\pi e - 0.0003 f_\pi e^3 .
\end{eqnarray}
Taking all this together, and subtracting three times the $B=1$ result,
 we find
\begin{equation}
E_{\rm gs} = -1.3 f_\pi/e + 0.781 f_\pi e -0.007 f_\pi e^3.
\end{equation}
If we use our three parameter sets, we find values
of $185.5~\rm MeV$ for ANW,  $186.5~\rm MeV$ for ANW' and 
$211.6~\rm MeV$ for R. These values are still high above the
threshold, but please note that we have overestimated the effect
of the lowest five modes by treating them as independent. Furthermore,
some of the higher modes lead to a separated single hedgehog and
a doughnut. Our experience from the $B=2$ case suggests that in such cases
the harmonic approximation may be relatively poor, and again leads
to an overestimate. We have found a reduction of the energy, however, which
is at least promising for a more mature approach to the problem.

\section{Conclusions and Outlook}
\label{sec:conclusions}

We have shown that our best treatment for the $B=2$ problem leads to
a state close to threshold (a few tens of MeV above or below) for
a reasonable choice of parameters. At the moment this uses a enhanced
version of the quantisation in the attractive manifold. What is really
missing is the Hamiltonian in the full manifold, which is not
readily accessible, even though one might consider cross-linking
the three channels discussed in our paper, Ref.~\cite{SkLACM}.
Using a different numerical approach to study the modes leading away
from the $B=2$ hedgehog, as done
by Waindzoch and Wambach \cite{WaWa}, may be another useful tool
to get more information. It is unlikely that any of these methods
will be able to completely cover the whole collective surface, however.

A more pressing problem is the issue whether we can really construct a 
decoupled 12-dimensional collective surface. Our results about level
crossings in the attractive channel seem to suggest that it may not be
possible to do so -- or maybe we can, but we need to introduce a 
monopole geometric phase in our collective Hamiltonian, which should influence
the dynamics and the rotational quantum numbers. 
This is an important issue in the quantisation of
the $B=2$ system.

The $B=3$ system apparently benefits a lot when the anharmonicities
are taking seriously. In the present calculation we have not treated 
all of them. That is at least partly due to the fact that the JNR
instantons are not the most general ones for this case. The fact
that the lowest two modes separate in two bananas instead of three
$B=1$ Skyrmions as the monopole result from Ref.\ \cite{HouSut} seems
to suggest should happen, is probably related to this deficiency.

It is therefore necessary to see whether one can handle the AHDM instanton
\cite{AHDM} in calculations of the type performed above. This is
not trivial, since the relation between Skyrmion and AHDM data
is rather indirect, and may just be too time consuming to implement.
We are currently examining the feasibility of this approach. If we can we will
be able to better address the anharmonicities for $B=3$ and probably
also see whether large amplitude fluctuations play a r\^ole for the $B=4$
system.

\section*{Acknowledgements}
I thank the Institute for Nuclear Theory at the University of Washington for
its hospitality and the Department of Energy for Partial support during
the initial stages of this work. I also acknowledge partial support by the 
Bundesministerium f{\"u}r Forschung und Technologie. I benefited from 
discussions with N. Manton and C.J. Houghton.

\appendix
\section{A detailed analysis of the modes of the tetrahedron.}
The instanton field is generated by four poles, arranged in an tetrahedron.
All weights are equal. We choose 
\begin{eqnarray}
(\lambda_1,X_1) &=& (1/4,-R,-R,-R,0),\nonumber\\
(\lambda_2,X_2) &=& (1/4,-R,R,R,0),\nonumber\\
(\lambda_3,X_3) &=& (1/4,R,-R,R,0),\nonumber\\
(\lambda_4,X_4) &=& (1/4,R,R,-R,0),
\end{eqnarray}
where $R$ is a parameter determining the size of the tetrahedron.

Since all continuous symmetries are broken, we find 9 zero modes.
In order of increasing frequency the non-zero 
modes can be interpreted as follows
(we concentrate on how the pole positions change. Of course the weights change
as well, but it is harder to classify modes that way.)
Since we discuss fluctuations here, they have a natural identification
as the gradients of a wave function. We shall identify those with the standard
molecular orbitals found e.g., in Refs.~\cite{Chen,Tinkham}. We shall also
apply the standard group representation labels corresponding to these
wave functions. 
\begin{itemize}
\item[10-11]
$\hbar\omega=0.318f_\pi e$, irrep=$E$.\\
The two eigenvectors can be chosen to be
\begin{eqnarray}
\delta(\lambda_1,X_1) &=& (0,-1,1,0,0),\nonumber\\
\delta(\lambda_2,X_2) &=& (0,-1,-1,0,0),\nonumber\\
\delta(\lambda_3,X_3) &=& (0,1,-1,0,0),\nonumber\\
\delta(\lambda_4,X_4) &=& (0,1,1,0,0),
\end{eqnarray}
and
\begin{eqnarray}
\delta(\lambda_1,X_1) &=& (0,-1,-1,2,0),\nonumber\\
\delta(\lambda_2,X_2) &=& (0,1,-1,-2,0),\nonumber\\
\delta(\lambda_3,X_3) &=& (0,-1,1,-2,0),\nonumber\\
\delta(\lambda_4,X_4) &=& (0,1,1,2,0).
\end{eqnarray}
They correspond to the gradient of the standard wave functions $x^2-y^2$ and
$x^2+y^2-2z^2$, respectively.
The two modes, both invariant
under a rotation around the x-axis, can be interpreted as follows.
One mode corresponds to rotating the
two lines (12) and (34) relative to one-another around the x-axis.
The other correspond to stretching the lines (12) and (34), 
at the same time moving the 
lines closer. Both would be expected to go to a planar configuration of 
poles arranged in a square, the $B=3$ doughnut.

\item[12-14] 
$\hbar\omega=0.332f_\pi e$, irrep=$T_2$.\\
One representative of this mode, the one
invariant under a  $120^\circ$ rotation about the 111 axis,
has the form
\begin{eqnarray}
\delta(\lambda_1,X_1) &=& (3\delta\lambda,1,1,1,0),\nonumber\\
\delta(\lambda_2,X_2) &=& (-\delta\lambda,-1,0,0,0),\nonumber\\
\delta(\lambda_3,X_3) &=& (-\delta\lambda,0,-1,0,0),\nonumber\\
\delta(\lambda_4,X_4) &=& (-\delta\lambda,0,0,-1,0)
\end{eqnarray}
The wave function
corresponding to the present mode is $-\half(yz + xz + xy)$, 
which has led to the $T_2$ assignment.

In this mode we always keep the three points 2,3,4 in a shrinking equilateral
triangle, at the same time moving pole 1 further and further
away. This should probably connect to a well separated system
consisting of the $B=2$ doughnut and a single hedgehog.

\item[15-17]
$\hbar\omega=0.469f_\pi e$, irrep=$T_2$.\\
One representative of this mode, the one
with eigenvalue 1 for a $120^\circ$ rotation around the 111 axis,
has the form
\begin{eqnarray}
\delta(\lambda_1,X_1) &=& (-3\delta\lambda,3,3,3,0),\nonumber\\
\delta(\lambda_2,X_2) &=& (\delta\lambda,-1,1,1,0),\nonumber\\
\delta(\lambda_3,X_3) &=& (\delta\lambda,1,-1,1,0),\nonumber\\
\delta(\lambda_4,X_4) &=& (\delta\lambda,1,1,-1,0)
\end{eqnarray}
The change in pole position is up to an overall translation in the $(1,1,1)$
direction the same as found above. The important difference is that the
size of $\delta\lambda$ is much larger here (0.274), whereas the value
found for the previous modes (.0383) is not incompatible with 0.

Another interpretation is a rescaling of all poles, 
with a factor $1+\alpha$ for poles 2,3,4 and $1-3\alpha$ for pole 1.

\item[18] 
$\hbar\omega=0.483f_\pi e$, irrep=$A_1$.\\
This is the breathing mode, as can easily be seen from the mode-vector.

\item[19-21]
$\hbar\omega=0.813f_\pi e$, irrep=$T_1$.\\
One representative of this mode, the one
with eigenvalue 1 for a $120^\circ$ rotation around the 111 axis,
has the form
\begin{eqnarray}
\delta(\lambda_1,X_1) &=& (0,0,0,0,3),\nonumber\\
\delta(\lambda_2,X_2) &=& (0,0,.6278,-.6278,1),\nonumber\\
\delta(\lambda_3,X_3) &=& (0,-.6278,0,.6278,1),\nonumber\\
\delta(\lambda_4,X_4) &=& (0,.6278,-.6278,0,1),\nonumber\\
\delta(c) &=& (1.178,1.178,1.178)
\end{eqnarray}
This is the only mode that mixes with iso-spin rotations. The change in
the space components is consistent with the irrep $T_1$ and 
a wave function $x(y^2-z^2)+y(z^2-x^2)+z(x^2-y^2)$.
After a study of the change in baryon number associated with this mode
we conclude that it probably corresponds to a twist, where we turn the
baryon density in the 234 triangle, keeping the edges fixed at the point
1. At the same time a complicated grooming of the triangle relative
to the pole 1 takes place (time-components). 
\end{itemize}

\end{document}